Superconducting gap of overdoped $Tl_2Ba_2CuO_{6+\delta}$ observed by Raman scattering

K.Nishikawa[1], T.Masui[1], S.Tajima[1], H.Eisaki[2], H.Kito[2], A.Iyo[2]

[1]Department of Physics, Graduate School of Science, Osaka University, Osaka 560-0043, Japan

[2]Nanoelectronics Research Institute, National Institute of Advanced Industrial Science and Technology (AIST), Tsukuba 305-8568, Japan

**Abstract**

We report Raman scattering spectra for single crystals of overdoped $Tl_2Ba_2CuO_{6+\delta}$ (Tl-2201) at low temperatures. It was observed that the pair-breaking peaks in $A_{1g}$ and $B_{1g}$ spectra radically shift to lower energy with carrier doping. We interpret it as $s$-wave component mixing into $d$-wave, although the crystal structure is tetragonal. Since similar phenomena were observed also in $YBa_2Cu_3O_y$ and $Bi_2Sr_2CaCu_2O_z$, we conclude that $s$-wave mixing is a common property for overdoped high-$T_c$ superconductors.

**Introduction**

Since the discovery of high-$T_c$ superconductors, the pairing mechanism is the most important problem. It is crucial for this study to know the symmetry of the order parameter. There have been a lot of experiments which addressed this problem. For hole doped high-$T_c$ superconductors at the optimal doping, it is established that the pairing states have the $d_{x^2-y^2}$ symmetry. But the symmetry of the order parameter appears to change with carrier doping. A possibility of *s*-wave mixing has been suggested by many researchers[1-7].

Electronic Raman scattering plays a special role in addressing this problem. For YBa$_2$Cu$_3$O$_{7-\delta}$(Y-123) in $A_{1g}$ and $B_{1g}$ polarizations the polarization dependence of pair-breaking peak energy disappeared in the very overdoped regime[2]. The change of peak energy in $B_{1g}$ polarization did not scale with the change of $T_c$. These observations indicate *s*-wave mixing into *d*-wave symmetry in the overdoped regime[1]. However, the symmetry of order parameter is generally affected by the symmetry of crystal structure. Especially for Y-123, orthorhombicity becomes larger with increasing oxygen content. Thus, *s*-wave mixing may be considered as a result of the increased orthorhombicity.

In order to address this problem, it is necessary to examine a tetragonal system. Since reports on Raman spectrum of Tl$_2$Ba$_2$CuO$_{6+\delta}$ (Tl-2201) are very limited[8,9], it must be helpful to study a precise doping dependence of Raman response in this material. In the present study, we prepared single crystals of tetragonal Tl-2201 and measured the Raman scattering spectra for three overdoped Tl-2201 crystals, focusing on the change of superconductivity response with respect to doping levels.

**Experiment**

Tl-2201 single crystals were grown by a flux method[10]. As-grown crystals were overdoped with a broad superconductivity transition, which suggests inhomogeneous oxygen distribution. Oxygen contents in the crystals were controlled by the annealing process in Ar atmosphere. All crystals were characterized by a SQUID magnetometer. $T_c$ values of three crystals were found to be 80K, 62K, and 52K, respectively. Since higher oxygen content gives lower $T_c$, all crystals are supposed to be overdoped.

We measured Raman scattering spectra by using T64000 Jobin-Yvon triple spectrometer with a liquid-nitrogen cooled CCD detector, and Ar-Kr laser. The wavelength of excited beam is 514.5nm. A closed-cycle cryostat was used with temperature stabilization better than 1K. The laser beam with about 4 mW power was focused into a spot of 50 $\mu$m diameter. Although it is not easy to cleave a crystal of Tl-2201, all the spectra were measured on fresh surfaces obtained by cracking crystals.

The crystal structure of Tl-2201 belongs to the tetragonal $D_{4h}$ point group. We extracted $B_{1g}(X'Y')$, $B_{2g}(XY)$, and the $A_{1g}(XX'-XY)$ scattering component from various polarization spectra. The $X'$ and $Y'$ axes are rotated by 45° with respect to the $X$ and $Y$ axes, respectively, which are parallel to the Cu-O bonds in the Tl-2201 unit cell. After subtraction of the dark counts of the detector the spectra were corrected for the Bose factor in order to obtain the imaginary part of the Raman response function.

**Results and Discussions**

The Raman spectra show some sharp phonon peaks and a broad electronic continuum. The superconducting transition causes the redistribution of the electronic continuum into a broad peak. Although not the focus of this study, there is a redistribution of phonon strength and positions. Figure 1 shows $B_{1g}$ and $A_{1g}$ Raman spectra of Tl-2201 single crystals above(black line) and well below(gray line) $T_c$. For $A_{1g}$ scattering in order to obtain the better view of the electronic continuum we subtract the spectra above $T_c$ from the spectra well below $T_c$ (Figure 2).

For the lightly overdoped crystal ($T_c$=80K) $B_{1g}$ and $A_{1g}$ pair-breaking peak intensities are very large. The pair-breaking peak energy ($E_p$) is dependent on the polarizations, as $E_p(B_{1g})\approx380$cm$^{-1}$ > $E_p(A_{1g})\approx300$cm$^{-1}$. For the most overdoped crystal ($T_c$=52K) $B_{1g}$ pair-breaking peak intensity is still large, but for $A_{1g}$ scattering it is largely reduced. Moreover, pair-breaking peak energy is independent of the polarizations, $E_p(B_{1g})\approx E_p(A_{1g})\approx140$cm$^{-1}$. For $B_{1g}$ scattering component, compared to the case of lightly overdoped crystal ($T_c$=80K) the pair breaking peak energy was reduced by half, although $T_c$ decreased only by 30%. If one identifies the peak in the $B_{1g}$ scattering component as $2\Delta_0$, one obtains the reduced gap value $2\Delta_0/k_BT_c\approx3.9$ for the most overdoped crystal, while in the lightly overdoped crystal it is about 6.8. This drastic reduction of the $B_{1g}$ pair-breaking peak energy takes place between the doping levels for the $T_c$=62K sample and the $T_c$=52K sample. A similar behavior was observed in $Bi_2Sr_2CaCu_2O_{8+\delta}$(Bi-2212) and Y-123 single crystals[2,11,12].

Several reasons can be considered for this disappearance of polarization dependence of the pair-breaking peak in very overdoped regime. The first possibility is that the symmetry of the order parameter changes from *d*-wave to an isotropic *s*-wave.

Although this interpretation was suggested by Kendziora *et al*.[8], it is unlikely because the low frequency profile of the $B_{1g}$ spectrum shows $\omega^3$-dependence, as is seen in the inset of Fig.1. This power law is in agreement with the theoretical prediction for *d*-wave superconductor [13], and in strong contrast to a sharp threshold excitation for *s*-wave gap. This suggests that the major component of the order parameter is *d*-wave symmetry with gap nodes.

The second possibility is that the Coulomb screening effects disappears in the overdoped regime. In the optimally doped regime, as it has *d*-wave symmetry, the Coulomb screening effect appears only in the $A_{1g}$ polarization. Therefore, $E_p(A_{1g})$ is reduced, which leads to the situation that $E_p(B_{1g}) > E_p(A_{1g})$. If this screening effects for $A_{1g}$ symmetry disappears, $E_p(A_{1g})$ is considered to be comparable to $E_p(B_{1g})$. Actually in the three-layer compounds such as Bi-2223, this behavior is observed due to the loss of screening effects [14]. However, since the $A_{1g}$ peak intensity is strongly suppressed in the heavily overdoped case, we conclude that the screening effect does exist in the present crystals.

As the last possibility, we need to think that the *s*-wave component is mixed into the *d*-wave symmetry. This was proposed by Nemetschek *et al* [1]. When *s*-wave component is mixing into *d*-wave, the order parameter is given by $\Delta(\mathbf{k})=\Delta_d[\cos(2\varphi)+r]$. Here $\varphi$ is the angle that $\mathbf{k}$ makes with the axis in the *a-b* plane and $r$ represents the *s*-wave mixing rate. In this case unscreened Raman response has two peaks at $2\Delta_d(1+r)$ and $2\Delta_d(1-r)$, and the screening effect is not only in the $A_{1g}$ polarization but also in the $B_{1g}$ polarization. When $r$ becomes larger, the higher peak energy in the $B_{1g}$ scattering is smeared out by the screening effect. Then the $B_{1g}$ pair-breaking peak energy corresponds to $2\Delta_d(1-r)$. This is smaller than the actual gap value $2\Delta_d$. According to this scenario, the disappearance of the polarization dependence of the peak energy for our very overdoped crystal($T_c$=52K) is due to screening effect in the $B_{1g}$ polarization spectrum as a result of *s*-wave mixing. The dramatic reduction of the peak energy is considered to be caused by the same reason.

According to Branch and Carbotte [15], $E_p(B_{1g})$ is sensitive to the change of Fermi surface topology in the anti-nodal direction. But the dramatic reduction of $E_p(B_{1g})$ observed in our result cannot be explained only by this effect. Therefore, we think $E_p(B_{1g})$ is reduced mainly due to *s*-wave mixing.

It should be noted that the *s*-wave mixing was observed in spite of the tetragonal structure of Tl-2201. This indicates that *s*-wave mixing is not due to the orthorhombicity which was discussed in Y-123 and Bi-2212, but it is a common property for overodped high-$T_c$ superconductors. The origin of the mixing of *s*-wave component is not clear yet.

For Y-123 single crystals, it is suggested that the dramatic reduction for $B_{1g}$ pair-breaking peak energy at the doping level p~0.19 may be related to a quantum critical point (QCP) [16]. If the present data of Tl-2201 are discussed from the same viewpoint, there may be a QCP between the moderate overdoping ($T_c$=62K) and the very overdoping ($T_c$=52K). According to the estimation of the hole concentration (*p*) by Tallon *et al* [17], $p \approx 0.22$ for $T_c$=62K and $p \approx 0.23$ for $T_c$=52K in Tl-2201, when $T_c$(max) is assumed to be 90K. Therefore, Tl-2201 may have a QCP at more overdoped state than the case for Y-123.

**Summary**

We prepared Tl-2201 single crystals and measured Raman spectra for $A_{1g}$ and $B_{1g}$ polarization. We observed the disappearance of the polarization dependence of the pair-breaking peak energy and drastic reduction of peak energy in heavily overdoped regime. This indicates that *s*-wave mixing in the overdoped regime is not due to the orthorhombicity, but characteristic for all high-$T_c$ superconductors. It was also found that the quantum critical point, if it exists, is located at *p*=0.22-23 in Tl-2201.

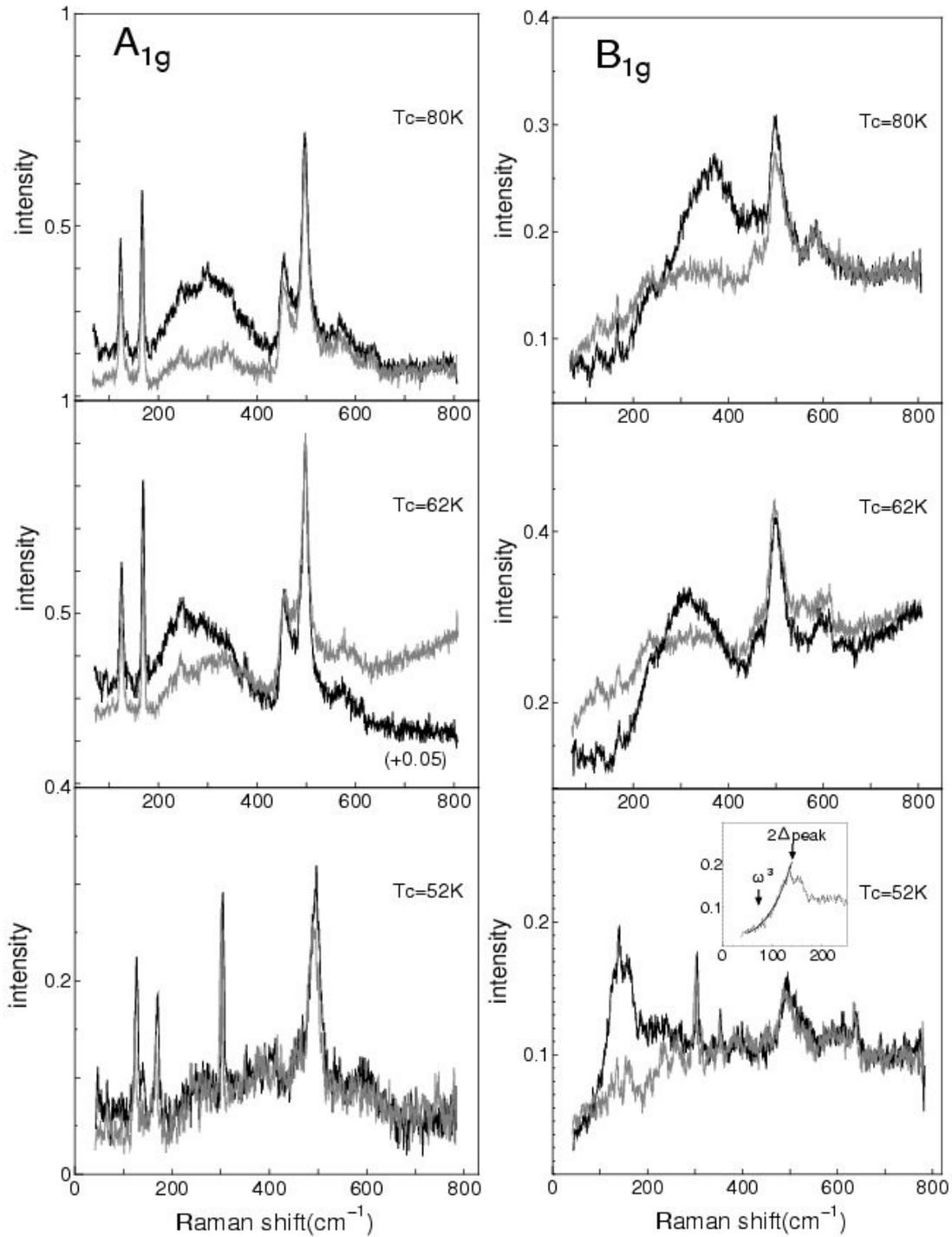

Fig. 1. Doping dependence of the Raman response of Tl-2201 single crystals for $A_{1g}$ (left panel) and $B_{1g}$ (right panel) polarizations. The columns are arranged from up to down in order of increasing doping level. The gray solid curves are the data taken above the respective $T_c$ of the samples. The black solid curves show the data taken in the SC state at $T \approx 10$K. Inset shows low temperature Raman response fitted by $\omega^3$ function.

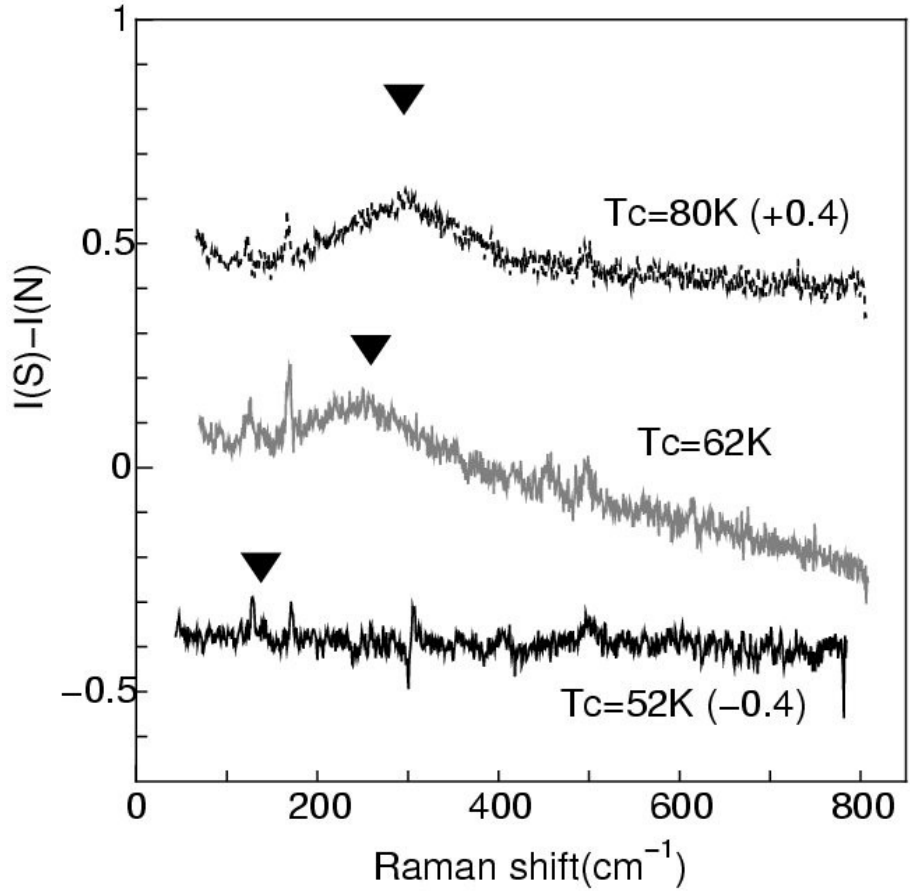

Fig. 2. $A_{1g}$ spectra where the normal state Raman response is subtracted from the SC state response. The spectra of $T_c$=80K and $T_c$=52K samples are shifted in the vertical scale by 0.4, respectively.

| $T_c$ | 80K | 62K | 52K |
|---|---|---|---|
| $2\Delta(A_{1g})$ | 300cm$^{-1}$ | 250cm$^{-1}$ | 140cm$^{-1}$ |
| $2\Delta(B_{1g})$ | 380cm$^{-1}$ | 310cm$^{-1}$ | 140cm$^{-1}$ |
| $2\Delta(B_{1g})/k_B T_c$ | 6.8 | 7.2 | 3.9 |

TABLE 1. Peak energies of the $A_{1g}$ and $B_{1g}$ electronic Raman scattering components for three samples.